\documentclass[aps,rpl,preprint,groupedaddress]{revtex4}
\begin{document}

\title{Local noise can enhance entanglement teleportation}
\author{Ye Yeo}
\affiliation{Department of Physics, National University of Singapore, 10 Kent Ridge Crescent, Singapore 119260, Singapore}

\begin{abstract}
Recently we have considered two-qubit teleportation via mixed states of four qubits and defined the generalized singlet fraction.  For single-qubit teleportation, Badziag {\em et al.} [Phys. Rev. A {\bf 62}, 012311 (2000)] and Bandyopadhyay [Phys. Rev. A {\bf 65}, 022302 (2002)] have obtained a family of entangled two-qubit mixed states whose teleportation fidelity can be enhanced by subjecting one of the qubits to dissipative interaction with the environment via an amplitude damping channel.  Here, we show that a dissipative interaction with the local environment via a pair of time-correlated amplitude damping channels can enhance fidelity of entanglement teleportation for a class of entangled four-qubit mixed states.  Interestingly, we find that this enhancement corresponds to an enhancement in the quantum discord for some states.  
\end{abstract}

\maketitle

Composite systems of two or more quantum objects $A, B, \cdots$ have interesting properties that are absent in quantum systems composed of single object.  Specifically, the principle of quantum superposition gives rise to the phenomenon of {\em entanglement} - a mysterious connection between separated quantum objects, which Einstein, Podolsky and Rosen \cite{Einstein} pointed out was a feature of quantum mechanics.  For two-qubit systems, the locally unitarily equivalent Bell basis states, $|\Psi^{\mu}_{\rm Bell}\rangle_{AB} = (u^{\mu}_A \otimes u^0_B)(|00\rangle_{AB} + |11\rangle_{AB})/\sqrt{2}$, are a class of maximally entangled states.  Here, $u^0$ is the $2 \times 2$ identity matrix; $u^1 = \sigma^1$, $u^2 = i\sigma^2$, $u^3 = \sigma^3$, and $\sigma^j$ ($j = 1, 2, 3$) are the Pauli matrices.

Interactions with the environment and imperfections of preparation result in noisy or mixed states described by density operators.  For instance, $|\Psi^{\mu}_{\rm Bell}\rangle_{AB}\langle\Psi^{\mu}_{\rm Bell}| \longrightarrow \chi_{AB} = {\cal E}(|\Psi^{\mu}_{\rm Bell}\rangle_{AB}\langle\Psi^{\mu}_{\rm Bell}|) = \sum_{\nu} E^{\nu}_{AB}|\Psi^{\mu}_{\rm Bell}\rangle_{AB}\langle\Psi^{\mu}_{\rm Bell}|E^{\nu\dagger}_{AB}$, where ${\cal E}$ is a quantum operation or channel ($E^{\nu}_{AB}$'s are the corresponding Kraus operators) \cite{Kraus}, which mathematically describes the noise and the resulting decoherence.  In general, the dissipative effects of noise degrade quantum entanglement.  A density operator $\rho_{AB}$ is separable if it can be written as a convex sum of separable pure states \cite{Werner}: $\rho_{AB} = \sum_k p_k |\psi^k\rangle_A\langle\psi^k| \otimes |\phi^k\rangle_B\langle\phi^k|$, where $\{p_k\}$ is a probability distribution and, $|\psi^k\rangle_A$ and $|\phi^k\rangle_B$ are vectors belonging to Hilbert spaces ${\cal H}_A$ and ${\cal H}_B$ respectively.  For two-level systems, a necessary and sufficient condition for separability is that a matrix, obtained by partial transposition of $\rho_{AB}$, has only non-negative eigenvalues \cite{Peres}.  In fact, as a measure of the amount of entanglement associated with a given two-qubit state $\rho_{AB}$, we may consider the negativity \cite{Vidal}: ${\cal N}[\rho_{AB}] \equiv \max\{0,\ -2\sum_m \lambda_m\}$, where $\lambda_m$ is a negative eigenvalue of $\rho^{T_B}_{AB}$, the partial transposition of $\rho_{AB}$.  Here, we focus on two-level systems.

Recent investigations have indicated that there are other properties associated with two-qubit states.  Groisman {\em et al.} \cite{Groisman} argued that $|\Psi^0_{\rm Bell}\rangle_{AB}$ contains one bit of ``quantum correlation'' and one bit of ``classical correlation''.  The total amount of correlation in a bipartite quantum state $\rho_{AB}$, ${\cal C}_{\rm total}[\rho_{AB}]$, is equal to the quantum mutual information $I(A:B) \equiv S[\rho_A] + S[\rho_B] - S[\rho_{AB}]$, where $\rho_A = {\rm tr}_B(\rho_{AB})$, $\rho_B = {\rm tr}_A(\rho_{AB})$, and von Neumann entropy $S[\rho] = -{\rm tr}[\rho\log_2\rho]$.  Clearly, for $|\Psi^0_{\rm Bell}\rangle_{AB}$, we have ${\cal C}_{\rm total}[|\Psi^0_{\rm Bell}\rangle_{AB}] = 2$.  To obtain the amount of classical correlation asociated with $|\Psi^0_{\rm Bell}\rangle_{AB}$, they determined ${\cal C}_{\rm total}[\sigma_{AB}]$, where $\sigma_{AB} = (|00\rangle_{AB}\langle 00| + |11\rangle_{AB}\langle 11|)/2$ is the state resulting from the erasure of the entanglement between $A$ and $B$.  That is, ${\cal C}_{\rm classical}[|\Psi^0_{\rm Bell}\rangle_{AB}] = {\cal C}_{\rm total}[\sigma_{AB}] = 1$; or ${\cal C}_{\rm quantum}[|\Psi^0_{\rm Bell}\rangle_{AB}] = {\cal C}_{\rm total}[|\Psi^0_{\rm Bell}\rangle_{AB}] - {\cal C}_{\rm classical}[|\Psi^0_{\rm Bell}\rangle_{AB}] = 1$.  In Ref.\cite{Ollivier}, Ollivier and Zurek introduced the quantum discord
\begin{equation}
{\cal D}_A(A:B) \equiv \sum^1_{m = 0}\pi_mS[\rho_{A|\Pi^m_B}]  + S[\rho_B] - S[\rho_{AB}],
\end{equation}
where the projectors $\Pi^m_B = |\pi^m\rangle_B\langle\pi^m|$ (with $|\pi^0\rangle \equiv \cos\theta|0\rangle + e^{i\phi}\sin\theta|1\rangle$, $|\pi^1\rangle = e^{-i\phi}\sin\theta|0\rangle - \cos\theta|1\rangle$, and $-\pi \leq \theta \leq \pi$, $0 \leq \phi \leq 2\pi$) describe perfect measurements of $B$; $\rho_{A|\Pi^m_B} = {\rm tr}_B(\Pi^m_B\rho_{AB}\Pi^m_B)/\pi_m$ is the state of $A$ after the measurement outcome $m$ has been detected; $S[\rho_{A|\Pi^m_B}]$ is the missing information about $A$, and probability $\pi_m = {\rm tr}[\Pi^m_B\rho_{AB}]$.  In general, this quantity depends both on $\rho_{AB}$ and $\{\Pi^m_B\}$, and is asymmetric under the change $A \leftrightarrow B$.  It is an information-theoretic measure of the quantum nature or ``quantumness'' of the correlations between $A$ and $B$.  Zurek \cite{Zurek} subsequently showed that quantum Maxwell's demons can extract more work than classical ones from correlations between a pair of quantum systems and that the difference is given by the discord.  We note that for density operator $\tau_{AB} = t_{00}|00\rangle_{AB}\langle 00| + t_{01}|00\rangle_{AB}\langle 11| + t_{10}|11\rangle_{AB}\langle 00| + t_{11}|11\rangle_{AB}\langle 11|$, we have the minimum discord
\begin{equation}
{\cal D}_{\min}[\tau_{AB}] = {\cal C}_{\rm quantum}[\tau_{AB}],
\end{equation}
when $|\pi^0\rangle = |0\rangle$ and $|\pi^1\rangle = |1\rangle$.

Many of the profound results in quantum information theory \cite{Nielsen} are impossible without the resource of entanglement.  For instance, it enables one to perform teleportation \cite{Bennett} - a way to send quantum information about object(s) to other (distant) object(s) using entanglement.  The spatially separated sender (Alice $\cal A$) and receiver (Bob $\cal B$) are only allowed to perform local quantum operations and communicate among themselves via a classical channel.  Teleportation can indeed serve as a fundamentally important operational test of not only the presence but also the quality of entanglement.  Popescu \cite{Popescu} had explored the different aspects of entanglement by analyzing the ``usefulness'' of Werner (channel) states \cite{Werner} for single-qubit teleportation.  He showed that there are Werner states, which do not violate any Bell type inequality, but still can be useful for teleportation.  In an equally interesting paper, Badziag {\em et al.} \cite{Badziag} presented a class of two-qubit entangled states, which may be made useful or ``more useful'' for single-qubit teleportation by subjecting one of the qubits to dissipative interaction with the environment via an amplitude damping channel.  Lee and Kim \cite{Lee} were the first to consider teleportation of two-qubit states via two independent, equally entangled Werner states.  In their scheme, Alice's joint measurement is decomposable into two independent Bell measurements and Bob's unitary recovery operation into two local one-qubit Pauli rotations - i.e., theirs is a straightforward generalization of the standard teleportation protocol ${\cal T}_0$ \cite{Bennett}.  They illustrated that entanglement of the two-qubit input state is lost during the teleportation even when the Werner states have nonzero entanglement, and in order to teleport any nonzero entanglement the channel states should possess a critical value of minimum entanglement.  Entanglement is fragile to teleport, teleporting entanglement demands more stringent conditions on the channel states.  It can thus reveal other aspects of the nature of entanglement and deserves more detailed studies.  In Refs.\cite{YeoI, YeoII}, we obtained some general results for two-qubit teleportation via four-qubit entangled states and introduced the concept of the {\em generalized singlet fraction}.

In this paper, we study the effects of local noise on the usefulness of a class of four-qubit entangled states $\Xi(\alpha, \beta)$ (Eq.(13)) for two-qubit teleportation.  To set the stage, we provide in the next paragraph, a brief introduction to singlet fraction and teleportation fidelity for single-qubit teleportation; and a summary of the results of Badziag {\em et al.}.  This is followed by a presentation of the relevant results from Refs.\cite{YeoI, YeoII}, after which we show that the corresponding generalized singlet fraction can be enhanced by subjecting Alice's qubits to dissipative interaction with the environment via a pair of time-correlated amplitude damping channels \cite{Yeo}.  In addition, we show that this enhancement corresponds to an enhancement in the quantum discord for some states.

When Alice and Bob share an arbitrary two-qubit mixed state $\chi_{AB}$ as a resource, ${\cal T}_0$ acts as a generalized depolarizing channel $\Lambda^{\chi, {\cal T}_0}_B$, with probabilities given by the maximally entangled components of the resource \cite{Bowen, Albeverio}: $\rho^{\rm out}_B \equiv \Lambda^{\chi, {\cal T}_0}_B(|\psi\rangle_B\langle\psi|) = \sum^3_{\mu = 0}\langle\Psi^{\mu}_{\rm Bell}|\chi|\Psi^{\mu}_{\rm Bell}\rangle \times u^{\mu\dagger}_B|\psi\rangle_B\langle\psi|u^{\mu}_B$.  Here, $|\psi\rangle_B = a_0|0\rangle_B + a_1|1\rangle_B$, with $a_0, a_1 \in {\cal C}^1$ and $|a_0|^2 + |a_1|^2 = 1$, is an arbitrary ``unknown'' (input) state of a qubit.  Consequently, at Bob's end, the teleported (output) state $\rho^{\rm out}_B$ can only be a distorted copy of the state $|\psi\rangle_A$ initially held by Alice.  The reliability for teleportation of a given channel state $\chi_{AB}$ is quantitatively measured by the teleportation fidelity,
\begin{eqnarray}
\Phi[\Lambda^{\chi, {\cal T}_0}_B] & \equiv & \int d\psi\ {_B}\langle\psi|\rho^{\rm out}_B|\psi\rangle_B \nonumber \\
& = & \frac{1}{3} + \frac{2}{3}{\cal F}[\chi],
\end{eqnarray}
where the singlet fraction
\begin{equation}
{\cal F}[\chi] \equiv \langle\Psi^0_{\rm Bell}|\chi|\Psi^0_{\rm Bell}\rangle.
\end{equation}
The maximum teleportation fidelity depends on the maximal singlet fraction \cite{Horodecki, Albeverio}: $\Phi[\Lambda^{\chi, {\cal T}_{\rm opt}}_B] = 1/3 + 2{\cal F}_{\max}[\chi]/3$, where ${\cal F}_{\max}[\chi] \equiv \max_u\langle\Psi^0_{\rm Bell}|(u^0 \otimes u)\chi(u^0 \otimes u^{\dagger})|\Psi^0_{\rm Bell}\rangle$.  The maximization is over the set of all unitary operations $u$ on ${\cal C}^2$.  Clearly, a necessary condition for faithful teleportation is that Alice and Bob share {\em a priori} a maximally entangled channel state.  In order to be useful for ${\cal T}_0$, $\chi_{AB}$ must have ${\cal F}_{\max}[\chi] > 1/2$ \cite{Popescu, Horodecki, Bose}.   If the channel state $\chi_{AB}$ is mixed too much (${\cal F}_{\max}[\chi] \leq 1/2$), it will not provide for any better teleportation fidelity than that of an ordinary classical communication protocol.  Now, consider the following one-parameter family of two-qubit mixed states:
\begin{equation}
\xi_{AB} \equiv \sum^1_{\nu = 0}(u^0_A \otimes K^{\nu}_B)|\Psi^0_{\rm Bell}\rangle_{AB}\langle\Psi^0_{\rm Bell}|
(u^0_A \otimes K^{\nu\dagger}_B),
\end{equation}
where
\begin{equation}
K^0 = \left(\begin{array}{cc} \sqrt{q} & 0 \\ 0 & 1 \end{array}\right),\
K^1 = \left(\begin{array}{cc} 0 & 0 \\ \sqrt{1 - q} & 0 \end{array}\right)
\end{equation}
with $0 \leq q \leq 1$, are Kraus operators that define an amplitude damping channel.  Hereafter, $|0\rangle$ and $|1\rangle$ denote the excited and ground states respectively.  The amplitude damping channel is characterized by the parameter $q$, with $1 - q$ denoting the dissipation strength when a qubit interacts with the environment via this channel.  Badziag {\em et al.} \cite{Badziag} and Bandyopadhyay \cite{Bandyopadhyay} showed that subjecting $\xi_{AB}$ to local noise at Alice's site:
\begin{equation}
\xi_{AB} \longrightarrow \xi'_{AB} = \sum^1_{\nu = 0} (K^{\nu}_A \otimes u^0_B)\xi_{AB}(K^{\nu\dagger}_A \otimes u^0_B)
\end{equation}
may improve the maximal singlet fraction.  That is, there exist values of $q$ such that though ${\cal F}_{\max}[\xi] < 1/2$ we can have ${\cal F}_{\max}[\xi'] > 1/2$, and also $1/2 < {\cal F}_{\max}[\xi] \leq {\cal F}_{\max}[\xi']$.  This is intriguing because the dissipative interaction with qubit $B$, which degrades entanglement in the first place is utilized to improve the quality of $\xi_{AB}$ by applying it to qubit $A$.  Bandyopadhyay reasoned qualitatively that given any mixed channel state $\chi_{AB}$, the corresponding maximal teleportation fidelity is determined by both the amount of entanglement ${\cal N}[\chi]$, and the ``classical correlations'' between Alice's qubit $A$ and Bob's qubit $B$; and since ${\cal N}[\chi]$ cannot be increased by Alice's local operations (in fact, $\cal N[\xi'] < \cal N[\xi]$), her action, Eq.(7), would only have enhanced the ``classical correlations''.  According to Bandyopadhyay, the enhancement in the maximal singlet fraction is thus due to improved ``classical correlations''.

In Ref.\cite{YeoI}, we gave an explicit protocol ${\cal E}_0$ for faithfully teleporting arbitrary two-qubit states, $|\Psi\rangle_{A_1A_2} = \sum^1_{i, j = 0}a_{ij}|ij\rangle_{A_1A_2}$ with $a_{ij} \in {\cal C}^1$ and $\sum^1_{i, j = 0}|a_{ij}|^2 = 1$, employing genuine four-qubit entangled states
\begin{equation}
|\Upsilon^{00}(\theta_{12}, \phi_{12})\rangle_{A_3A_4B_1B_2} \equiv 
\frac{1}{2}\sum^3_{J = 0}|J\rangle_{A_3A_4} \otimes |J'\rangle_{B_1B_2}.
\end{equation}
$\{|J\rangle = S|ij\rangle\}$ and $\{|J'\rangle = T|ij\rangle\}$ are orthonormal bases, with
\begin{eqnarray}
S(\theta_1, \phi_1) & \equiv & \left(\begin{array}{cccc}
\cos\theta_1 & 0 & 0 & -\sin\theta_1 \\
0 & \cos\phi_1 & -\sin\phi_1 & 0 \\
0 & \sin\phi_1 & \cos\phi_1 & 0 \\
\sin\theta_1 & 0 & 0 & \cos\theta_1
\end{array}\right), \nonumber \\
T(\theta_2, \phi_2) & \equiv & \left(\begin{array}{cccc}
\cos\theta_2 & 0 & 0 & -\sin\theta_2 \\
0 & \sin\phi_2 & \cos\phi_2 & 0 \\
0 & \cos\phi_2 & -\sin\phi_2 & 0 \\
\sin\theta_2 & 0 & 0 & \cos\theta_2
\end{array}\right).
\end{eqnarray}
Here, $-\pi/2 < \theta_{12} \equiv \theta_1 - \theta_2 < \pi/2$ and $-\pi/2 < \phi_{12} \equiv \phi_1 - \phi_2 < \pi/2$, since $0 < \theta_1,\ \theta_2,\ \phi_1,\ \phi_2 < \pi/2$.  Whenever $\theta_{12} = \phi_{12} = 0$, $|\Upsilon^{00}\rangle$ is reducible to a tensor product of two Bell states: $|\Upsilon^{00}\rangle_{A_3A_4B_1B_2} = |\Psi^0_{\rm Bell}\rangle_{A_3B_2} \otimes |\Psi^0_{\rm Bell}\rangle_{A_4B_1}$.  Alice performs a complete projective measurement jointly on $A_1A_2A_3A_4$ in the following basis of 16 orthonormal states:
\begin{equation}
|\Pi^{\mu\nu}(\theta_{12}, \phi_{12})\rangle_{A_1A_2A_3A_4} 
\equiv (U^{\mu\nu}_{A_1A_2} \otimes U^{00}_{A_3A_4})|\Pi^{00}(\theta_{12}, \phi_{12})\rangle_{A_1A_2A_3A_4},
\end{equation}
with $|\Pi^{00}(\theta_{12}, \phi_{12})\rangle_{A_1A_2A_3A_4} \equiv \frac{1}{2}\sum^3_{K = 0}|K'\rangle_{A_1A_2} \otimes |K\rangle_{A_3A_4}$ and $U^{\mu\nu} \equiv u^{\mu} \otimes u^{\nu}$.  Upon receiving classical information of her measurement result, Bob can always succeed in recovering an exact replica of the original state of Alice's particles $A_1A_2$, by applying the appropriate recovery unitary operations to his particles $B_1B_2$.  If Alice and Bob share {\em a priori} two pairs of particles, $A_3A_4$ and $B_1B_2$, in an arbitrary four-qubit mixed state $\Xi_{A_3A_4B_1B_2}$ as a resource, ${\cal E}_0$ acts as a generalized depolarizing bichannel \cite{YeoII}: $\Lambda^{\Xi, {{\cal E}_0}}_{B_1B_2}(|\Psi\rangle_{B_1B_2}\langle\Psi|) = \sum^3_{\mu, \nu = 0}\langle\Upsilon^{\mu\nu}|\Xi|\Upsilon^{\mu\nu}\rangle \times U^{\mu\nu\dagger}_{B_1B_2}|\Psi\rangle_{B_1B_2}\langle\Psi|U^{\mu\nu}_{B_1B_2}$, where we define $|\Upsilon^{\mu\nu}\rangle \equiv (U^{00} \otimes U^{\mu\nu\dagger})|\Upsilon^{00}\rangle$.  The fidelity of teleportation
\begin{eqnarray}
\Phi[\Lambda^{\Xi, {\cal E}_0}_{B_1B_2}] & \equiv & \int d\Psi\ 
{_{B_1B_2}}\langle\Psi|\Lambda^{\Xi, {\cal E}_0}_{B_1B_2}(|\Psi\rangle_{B_1B_2}\langle\Psi|)|\Psi\rangle_{B_1B_2} \nonumber \\
& = & \frac{1}{5} + \frac{4}{5}{\cal G}[\Xi],
\end{eqnarray}
where the generalized singlet fraction
\begin{equation}
{\cal G}[\Xi] \equiv \max_{\theta_{12}, \phi_{12}}
\{\langle\Upsilon^{00}(\theta_{12}, \phi_{12})|\Xi|\Upsilon^{00}(\theta_{12}, \phi_{12})\rangle\},
\end{equation}
in contrast to Eqs.(3) and (4).  We note that the $\theta_{12}$ and $\phi_{12}$, which give ${\cal G}[\Xi]$, determine Alice's measurement, Eq.(10).  $\Xi$ is useful for ${\cal E}_0$ if ${\cal G}[\Xi] > 1/2$ and $\Phi[\Lambda^{\Xi, {\cal E}_0}_{B_1B_2}] > 3/5$.

Now we are ready to present our results.  Consider the four-qubit state
\begin{equation}
\Xi_{A_1A_2B_1B_2}(\alpha, \beta) = \sum^1_{\nu = 0} (U^{00}_{A_1A_2} \otimes K^{\nu\nu}_{B_1B_2})
|\Upsilon^{00}(\alpha, \beta)\rangle_{A_1A_2B_1B_2}\langle\Upsilon^{00}(\alpha, \beta)|
(U^{00}_{A_1A_2} \otimes K^{\nu\nu\dagger}_{B_1B_2}),
\end{equation}
which can be obtained in the following way: Alice prepares the four-qubit state $|\Upsilon^{00}(\alpha, \beta)\rangle$ (Eq.(8)) locally in her laboratory and sends any two qubits to Bob simultaneously across a pair of time-correlated amplitude damping channels  described by the Kraus operators \cite{Yeo}
\begin{equation}
K^{00} = \left(\begin{array}{cccc}
\sqrt{q} & 0 & 0 & 0 \\
0 & 1 & 0 & 0 \\
0 & 0 & 1 & 0 \\
0 & 0 & 0 & 1
\end{array}\right),\
K^{11} = \left(\begin{array}{cccc}
0 & 0 & 0 & 0 \\
0 & 0 & 0 & 0 \\
0 & 0 & 0 & 0 \\
\sqrt{1 - q} & 0 & 0 & 0
\end{array}\right).
\end{equation}
Its generalized singlet fraction is independent of both $\alpha$ and $\beta$, and is a simple function of $q$ given by
\begin{equation}
{\cal G}[\Xi(\alpha, \beta)] = \frac{1}{16}(3 + \sqrt{q})^2,
\end{equation}
when $\theta_{12} = \alpha$ and $\phi_{12} = \beta$.  Applying the prescription, similar to that in Ref.\cite{Bandyopadhyay}: Alice allows her pair of qubits $A_1$ and $A_2$ to interact with the environment via a pair of time-correlated amplitude damping channels of the same strength as above; we obtain
\begin{eqnarray}
\Xi_{A_1A_2B_1B_2}(\alpha, \beta) & \longrightarrow  & \Xi'_{A_1A_2B_1B_2}(\alpha, \beta) \nonumber \\
& = & \sum^1_{\nu = 0} (K^{\nu\nu}_{A_1A_2} \otimes U^{00}_{B_1B_2})\Xi_{A_1A_2B_1B_2}(\alpha, \beta)
(K^{\nu\nu\dagger}_{A_1A_2} \otimes U^{00}_{B_1B_2}).
\end{eqnarray}
To determine the corresponding generalized singlet fraction, we have $\phi_{12} = \beta$ but $\theta_{12}$ is in general a very complicated function of both $\alpha$ and $q$.  However, for $\alpha = \beta = 0$, we have
\begin{equation}
{\cal G}[\Xi'(0, 0)] = \frac{1}{8}(5 + 2q + q^2),
\end{equation}
when $\theta_{12} = \phi_{12} = 0$.  Both ${\cal G}[\Xi(0, 0)]$ and ${\cal G}[\Xi'(0, 0)]$ are strictly greater than $1/2$, and we have ${\cal G}[\Xi'(0, 0)] \geq {\cal G}[\Xi(0, 0)]$ if $0 < q \leq q_{\rm crit} \approx 0.0338454$.  The range of values of $q$ for which ${\cal G}[\Xi'(\alpha, \beta)] \geq {\cal G}[\Xi(\alpha, \beta)]$ shrinks as $\alpha$ differs more and more from zero.  For instance, when $\alpha = 0.1\pi$, we have $q_{\rm crit} \approx 0.0209421$.  An interesting question is what exactly does this enhancement in generalized singlet fraction physically correspond to?

In order to answer the above question, we consider input states $|\Psi\rangle_{B_1B_2} = \cos\epsilon |00\rangle + \sin\epsilon|11\rangle$ with $0 \leq \epsilon \leq \pi/4$.  For $\Xi(\alpha, \beta)$, the output states are $\Lambda^{\Xi, {\cal E}_0}_{B_1B_2}(|\Psi\rangle_{B_1B_2}\langle\Psi|) = \tau_{B_1B_2}$, with $t_{00} = \gamma_+$, $t_{01} = t_{10} = 1/8[2 + \sqrt{q} + q + (\sqrt{q} - q)\cos4\alpha]\sin2\epsilon$, $t_{11} = \gamma_-$, and $\gamma_{\pm} = \{4 \pm [2 + \sqrt{q} + q - (\sqrt{q} - q)\cos4\alpha]\cos2\epsilon\}/8$.  Straightforward calculations yield
\begin{equation}
{\cal N}[\Lambda^{\Xi, {\cal E}_0}_{B_1B_2}(|\Psi\rangle_{B_1B_2}\langle\Psi|)] = 
\frac{1}{4}[2 + \sqrt{q} + q + (\sqrt{q} - q)\cos4\alpha]\sin2\epsilon,
\end{equation}
which is approximately $0.550976\sin2\epsilon$ if $\alpha = 0.1\pi$ and $q = q_{\rm crit} \approx 0.0209421$.  For $\Xi'(0.1\pi, \beta)$, we obtain $\Lambda^{\Xi', {\cal E}_0}_{B_1B_2}(|\Psi\rangle_{B_1B_2}\langle\Psi|) = \tau_{B_1B_2}$, but with $t_{00} \approx 0.988715\cos^2\epsilon + 0.0112853\sin^2\epsilon$, $t_{01} = t_{10} \approx 0.508517\cos\epsilon\sin\epsilon$, $t_{11} \approx 0.988715\sin^2\epsilon + 0.0112853\cos^2\epsilon$, and ${\cal N}[\Lambda^{\Xi', {\cal E}_0}_{B_1B_2}(|\Psi\rangle_{B_1B_2}\langle\Psi|)] \approx 0.508517\sin2\epsilon$, which is smaller than ${\cal N}[\Lambda^{\Xi, {\cal E}_0}_{B_1B_2}(|\Psi\rangle_{B_1B_2}\langle\Psi|)]$.  In general, we have ${\cal N}[\Lambda^{\Xi', {\cal E}_0}_{B_1B_2}(|\Psi\rangle_{B_1B_2}\langle\Psi|)] \leq {\cal N}[\Lambda^{\Xi, {\cal E}_0}_{B_1B_2}(|\Psi\rangle_{B_1B_2}\langle\Psi|)]$.  This is not unexpected since with the addition of further noise to the channel state $\Xi$, the resulting generalized depolarizing bichannel $\Lambda^{\Xi', {\cal E}_0}_{B_1B_2}$ becomes more noisy, which degrades entanglement more.  One may conclude that, as in the case of single-qubit teleportation, the enhancement in the generalized singlet fraction is due to an improvement in the ``classical correlations'' and hence would not bring about an enhancement in any quantum property, such as entanglement, of the output states.  Surprisingly, we can show that there is enhancement in the quantum discord (Eq.(2)) for some output states whenever there is an enhancement in the generalized singlet fraction.  To this end, we calculate
\begin{equation}
{\cal D}_{\min}[\Lambda^{\Xi, {\cal E}_0}_{B_1B_2}(|\Psi\rangle_{B_1B_2}\langle\Psi|)] = 
-\gamma_+\log_2\gamma_+ - \gamma_-\log_2\gamma_- + \Gamma_-\log_2\Gamma_- + \Gamma_+\log_2\Gamma_+,
\end{equation}
where $\Gamma_{\pm} = 1/2 \pm \sqrt{2}/16 [8 + 8\sqrt{q} + 11q + 2q^{\frac{3}{2}} + 3q^2 + (1 - \sqrt{q})^2q\cos8\alpha - 4(2\sqrt{q} - q - q^2)\cos4\alpha\cos4\epsilon]^{\frac{1}{2}}$.  For definiteness, we consider $\alpha = 0.1\pi$ and $q = 0.01 < q_{\rm crit}$.  Straightforward calculations then yield ${\cal D}_{\min}[\Lambda^{\Xi', {\cal E}_0}_{B_1B_2}(|\Psi\rangle_{B_1B_2}\langle\Psi|)]$ as a function of $\epsilon$:
\begin{equation}
{\cal D}_{\min}[\Lambda^{\Xi', {\cal E}_0}_{B_1B_2}(|\Psi\rangle_{B_1B_2}\langle\Psi|)] = 
-\lambda_+\log_2\lambda_+ - \lambda_-\log_2\lambda_- + \Lambda_-\log_2\Lambda_- + \Lambda_+\log_2\Lambda_+,
\end{equation}
where
\begin{eqnarray}
\lambda_{\pm} & \approx & 0.5 \pm 0.00272371\sqrt{20765.4 + 12203.4\cos4\epsilon}, \nonumber \\
\Lambda_+ & \approx & 0.994553\cos^2\epsilon + 0.00544741\sin^2\epsilon, \nonumber \\
\Lambda_- & \approx & 0.994553\sin^2\epsilon + 0.00544741\cos^2\epsilon.\nonumber
\end{eqnarray}
Obviously, for $0 < \epsilon < 0.459496$, we have
$$
{\cal D}_{\min}[\Lambda^{\Xi', {\cal E}_0}_{B_1B_2}(|\Psi\rangle_{B_1B_2}\langle\Psi|)] >
{\cal D}_{\min}[\Lambda^{\Xi, {\cal E}_0}_{B_1B_2}(|\Psi\rangle_{B_1B_2}\langle\Psi|)].
$$

In conclusion, we have shown that a dissipative interaction with the local environment via a pair of time-correlated amplitude damping channels can enhance the generalized singlet fraction for a class of entangled four-qubit mixed states.  We demonstrate that this enhancement should correspond to an improvement in some quantum property of the four-qubit state by showing that the quantum discord for some output states is enhanced in the process.  It is hoped that the results will lead to a better understanding of multipartite entanglement.

\end{document}